\documentclass[10pt,letterpaper]{article}
\usepackage[top=0.85in,left=2.75in,footskip=0.75in]{geometry}

\usepackage{amsmath,amssymb}

\usepackage{changepage}

\usepackage[utf8x]{inputenc}

\usepackage{textcomp,marvosym}

\usepackage{cite}

\usepackage{nameref,hyperref}

\usepackage[right]{lineno}

\usepackage{microtype}
\DisableLigatures[f]{encoding = *, family = * }

\usepackage[table]{xcolor}

\usepackage{array}

\newcolumntype{+}{!{\vrule width 2pt}}

\newlength\savedwidth

\raggedright
\usepackage[aboveskip=1pt,labelfont=bf,labelsep=period,justification=raggedright,singlelinecheck=off]{caption}

\makeatletter
\renewcommand{\@biblabel}[1]{\quad#1.}
\makeatother

\usepackage{graphicx}
\usepackage{comment}
\usepackage{bm}
\usepackage{booktabs}
\usepackage{algorithm}
\usepackage{algorithmic}
\usepackage{multirow}
\usepackage{amsmath}
\usepackage{amsfonts}
\usepackage{url}
\usepackage{color}
\usepackage{enumerate}

\usepackage{lastpage,fancyhdr,graphicx}
\usepackage{epstopdf}
\fancyheadoffset[L]{2.25in}
\fancyfootoffset[L]{2.25in}
\begin{document}
\vspace*{0.2in}
\nolinenumbers

\begin{flushleft}
{\Large
\textbf\newline{Crowdsourced Hypothesis Generation and their Verification: A Case Study on Sleep Quality Improvement 
}}
\newline
\\
Shoko Wakamiya\textsuperscript{1},
Toshiki Mera\textsuperscript{2},
Eiji Aramaki\textsuperscript{1},
Masaki Matsubara\textsuperscript{2},
Atsuyuki Morishima\textsuperscript{2*}
\\
\bigskip
\textbf{1} Nara Institute of Science and Technology, Japan
\\
\textbf{2} University of Tsukuba, Japan
\\
\bigskip

%
%





* morishima-office@ml.cc.tsukuba.ac.jp

\end{flushleft}

\section*{Abstract}
A clinical study is often necessary for exploring important research questions; however, this approach is sometimes time and money consuming. Another extreme approach, which is to collect and aggregate opinions from crowds, provides a result drawn from the crowds' past experiences and knowledge. To explore a solution that takes advantage of both the rigid clinical approach and the crowds' opinion-based approach, we design a framework that exploits crowdsourcing as a part of the research process, whereby crowd workers serve as if they were a scientist conducting a ``pseudo" prospective study.
This study evaluates the feasibility of the proposed framework to generate hypotheses on a specified topic and verify them in the real world by employing many crowd workers. 
The framework comprises two phases of crowd-based workflow. In Phase 1—the hypothesis generation and ranking phase—our system asks workers two types of questions to collect a number of hypotheses and rank them. In Phase 2—the hypothesis verification phase—the system asks workers to verify the top-ranked hypotheses from Phase 1 by implementing one of them in real life. Through experiments, we explore the potential and limitations of the framework to generate and evaluate hypotheses about the factors that result in a good night's sleep.
Our results on significant sleep quality improvement show the basic feasibility of our framework, suggesting that crowd-based research is compatible with experts' knowledge in a certain domain. 



\section*{Introduction}
General research is consisting of two parts: to generate a new hypothesis and to validate it. Thus, to make a pool of potentially valid hypotheses is a key clue of researches. However, there are two difficult problems in the research process. First, to come up with a new hypothesis is a tough task, even for an experienced researcher. The second issue is to validate the hypothesis requires a high cost. In particular, clinical research handling human beings often incurs considerable costs in terms of the time required to conceive of approaches for a specified question and the monetary costs in terms of recruiting people. 

One of the solutions is to employ collective crowd intelligence, which could be used for both a hypothesis suggestion and validation, such as majority voting. 
So far, many attempts at citizen science involving crowdsourcing have been explored~\cite{franzoni2014,nov2011,ramine2010,vaish2010}.
In medical research, crowdsourcing has mainly been used to collect people's health data~\cite{swan2012,leiter2014}, questions and hypotheses~\cite{McAndrew2017,Crangle2018}. 
Also, there are attempts to obtain data from the crowd to support hypotheses on health ~\cite{bevelander2014} or to collect hypotheses by explicitly asking the crowd~\cite{aramaki2017}. 
However, it is known that results from such approaches using crowdsourcing are not always reliable~\cite{strickland2018} as they aggregate opinions drawn from the limited past experiences and knowledge of the crowd. Furthermore, even if hypotheses were obtained from the crowd, experts were eventually required to extract hypotheses that seem to be worth testing and actually verify some of them.

\begin{figure}[t]
 \centering
 \includegraphics[clip,width=10cm]{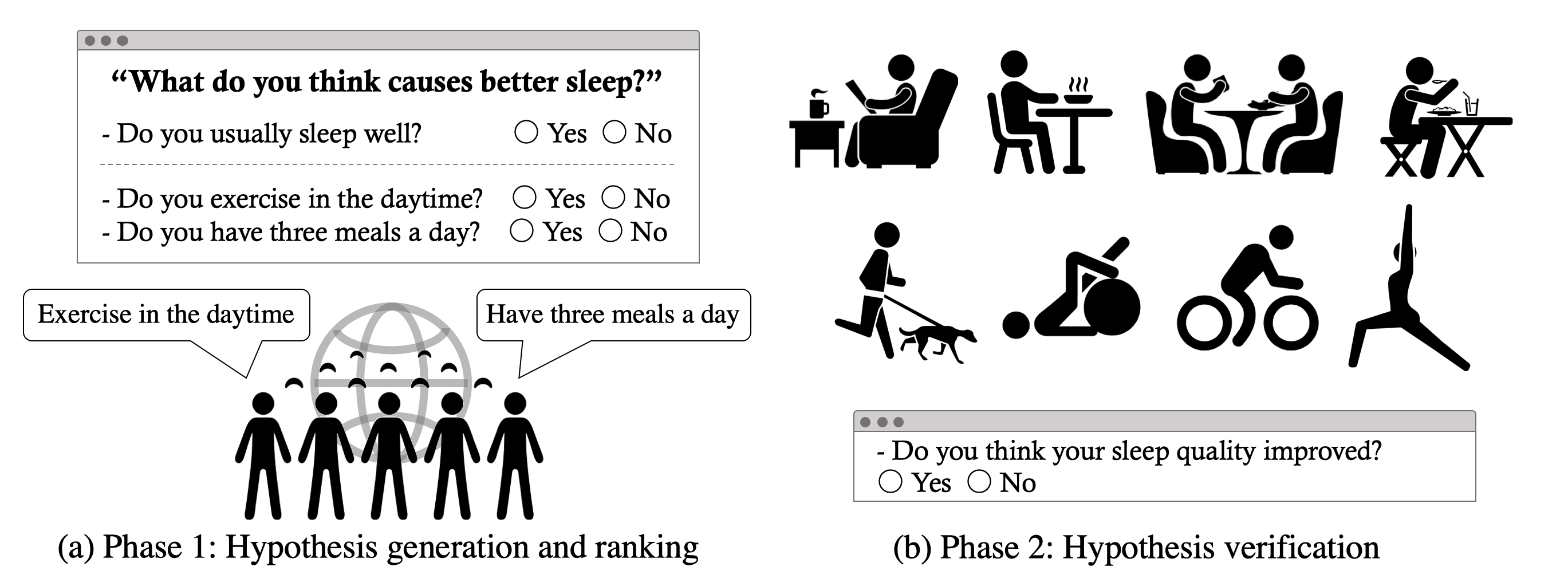}
 \caption{The framework involving in crowd workers consists of two workflow phases: the hypothesis generation and ranking phase and the hypothesis verification phase. (a) In Phase 1, crowd workers are asked to provide their hypotheses on a specified topic and rank them. (b) In Phase 2, the top-ranked hypotheses from Phase 1 are tested by the crowd workers in the real world. }
 \label{fig1}
\end{figure}

This paper aims to report our experience to explore the feasibility of a solution framework that takes advantage of both the rigid clinical approach and the crowd-based approach. Our framework exploits crowdsourcing as a part of the research process, whereby crowd workers (or simply workers) serve as if they were a scientist conducting a ``pseudo" prospective study. The workflow in the framework comprises two phases, as shown in Fig~\ref{fig1}, self-contained small tasks are distributed to a large number of workers to generate hypotheses for a given phenomenon. Then, these hypotheses are verified by workers to find possible causes. 

In Phase 1—the hypothesis generation and ranking phase—we collect a number of hypotheses from workers and rank the collected hypotheses using two types of questions. In particular, our system asks the workers to provide their hypotheses on a specified topic. For example, if the system asked the question ``What do you think causes better sleep?" the workers would provide their answers to this question. The system also shows a list of the top-ranked hypotheses, as voted by other workers, and asks the workers to judge whether they consider the hypotheses in the list to be true or false. The combination of these two types of questions results in a collection of hypotheses ranked in the order of potential feasibility. 

In Phase 2—the hypothesis verification phase—the system asks workers to verify the top-ranked hypotheses from Phase 1 by implementing one of them in real life. In other words, the workers themselves voluntarily serve as a small clinical trial. For example, if the hypothesis was ``Having three meals a day results in better sleep," the workers would eat three meals per day for a week in real life. After this trial period, the system asks the workers to report their trial results, from which we select worthwhile hypotheses. 

Our interest is the feasibility of the proposed framework, consisting of the two workflow phases; hypothesis producing (Phase 1) and its evaluation (Phase 2). We believe that the framework can also be used as a pre-study tool to identify hypotheses that deserve attention.

\section*{Methods}
This section explains the workflow design of the framework, which can be used to obtain and verify hypotheses on phenomena related to health conditions (called an outcome in medical statistics).  
The workflow consists of two phases; Phase 1 extracts a small number of hypotheses collected from crowd workers, and Phase 2 attempts to verify them by crowd workers. 

\subsection*{Phase 1: Generate and Rank Hypotheses}
In Phase 1, workers generate a ranked hypotheses list [$h_1$, ..., $h_m$] for the causes of an outcome. The ordering represents how likely they are to be true. First, the number of hypotheses increases linearly, and it is not easy to determine when to stop collecting hypotheses and start the selection. We solve this problem by designing a task to implement pipeline execution; the task contains questions for both collecting hypotheses and selecting them according to experience. Therefore, we are not concerned with when to start the selection process. 

\subsubsection*{Questions to be Asked}
Three types of questions are shown to each worker in Phase 1:

\begin{itemize}
    \item Outcome question $Q_{outcome}$: This question asks whether each worker has ever experienced a particular outcome. Let's say the outcome is ``Getting better sleep,'' the question may be direct, such as ``Do you usually sleep well?," or implicit, such as those in the Pittsburgh Sleep Quality Index (PSQI)~\cite{psqi}.
    \item Open hypothesis question $Q_{open}$: This question directly asks the workers for a hypothesis on the cause of A. If A is ``Getting better sleep," the question will be ``What do you think is a possible cause of good sleep?" The list of hypotheses generated by other workers can be displayed to reduce duplicates. The list can be given in a tree form.
    \item Closed hypothesis question $Q_{closed}$: This question shows the list of possible causes extracted from other workers' answers to the open hypothesis question $Q_{open}$, and it asks the worker whether the respective causes are consistent with the worker's past experience.
\end{itemize}

\begin{figure}[t]
 \centering
 \includegraphics[clip,width=10cm]{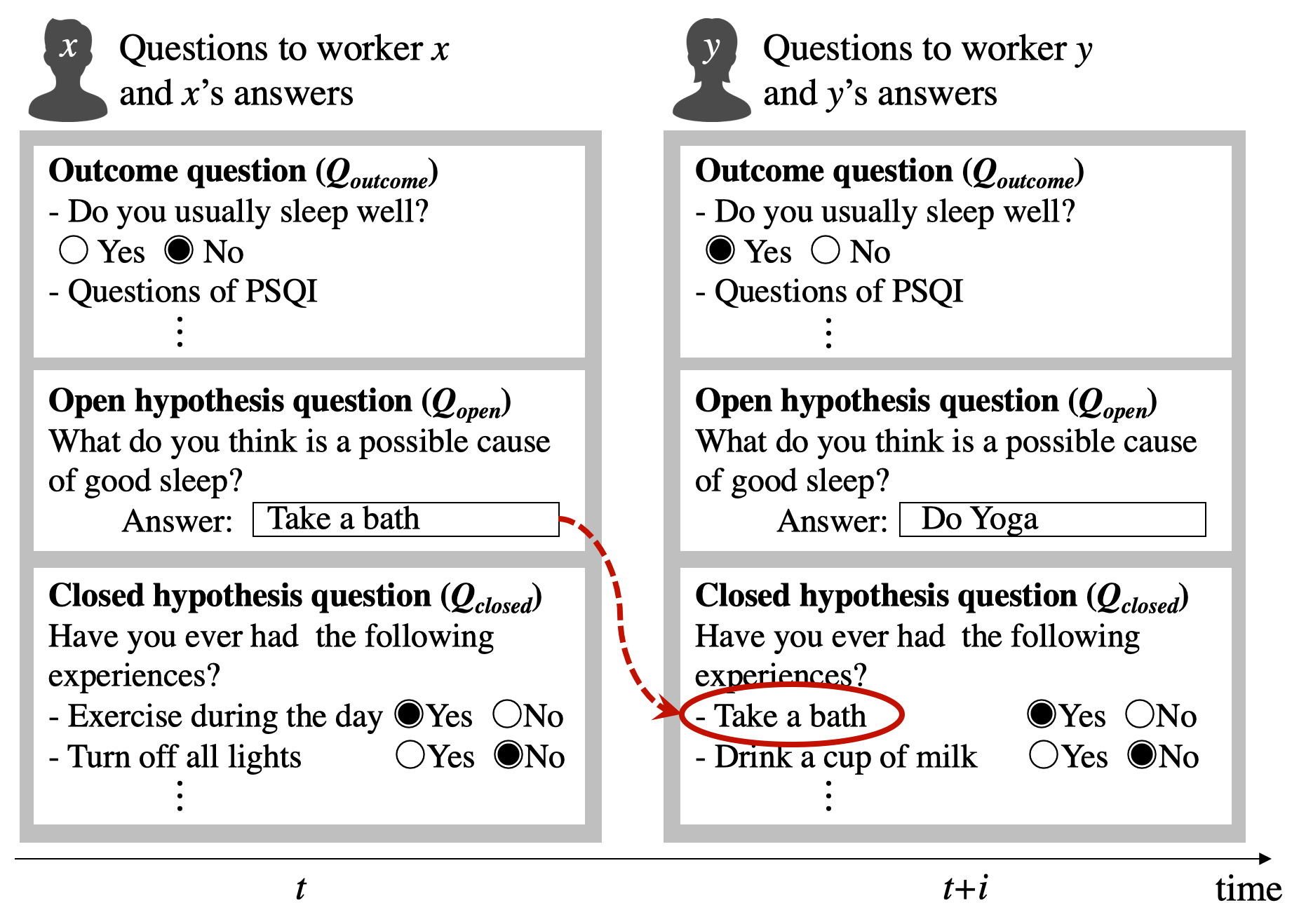}
 \caption{Phase 1: Pipeline execution for hypothesis generation and ranking. Outcome question $Q_{outcom}$ is to determine a worker's condition, open hypothesis question $Q_{open}$ is to obtain a hypothesis that could conduce to outcome, and closed hypothesis question $Q_{closed}$ is to rank possible causes obtained from other workers' answers to $Q_{open}$, e.g., $x$'s answer in $Q_{open}$ is shown in $y$'s tasks.}
 \label{fig2}
\end{figure}

\subsubsection*{Ranking}
Assuming that we do not have to limit the number of task submissions and workers and that we have sufficient workers to answer outcome and closed hypothesis questions, then it is possible to generate cross-tabulations with meaningful numbers. Now, it remains unclear as to which hypotheses deserve to be verified in Phase 2. 

A common approach in the field of medical research is to employ the odds ratio~\cite{aramaki2017,bevelander2014}, which compares the outcome in two groups (those who have experienced the possible cause and those who have never experienced it). Given a cross-tabulation for $h_i$ (Table~\ref{table1}), the odds ratio of hypothesis $h_i$ is computed as $P(h_i)= (a/b)/(c/d)$. This value represents how strongly the presence or absence of the possible cause is associated with the presence or absence of the outcome. Higher values mean that the possible cause mentioned in the hypothesis is more likely to be an actual cause. 

\begin{table}
\centering
\caption{Cross-tabulation used to calculate the hypothesis probability $P(h_i)$.}
\begin{tabular}{llc|c} \toprule
 & & \multicolumn{2}{c}{Worker's condition}\\ \cline{3-4}
 & & \multicolumn{1}{c|}{Sleep well} & Not sleep well\\ \hline
Consistency between worker's & \multicolumn{1}{|c|}{Yes} & $a$ & $c$\\ \cline{2-4}
experience and $h_i$ & \multicolumn{1}{|c|}{No} & $b$ & $d$\\ \bottomrule
\end{tabular}
\label{table1}
\end{table}

\subsubsection*{Combining Three Questions for Pipeline Execution}
A naive implementation of questionnaires involves two tasks. First, we use the open hypothesis questions to generate hypotheses, and then use the outcome and closed hypothesis questions to make cross-tables and rank the hypotheses. However, this would create another problem: knowing when to stop submitting hypotheses during the first task. Further, it is not easy to determine when to start filtering the collected hypothesis.

We solve this problem by combining all questions within the same task, named the Hypothesis Generation and Ranking Task (Fig~\ref{fig2}). For each task, the closed hypothesis questions are dynamically generated from the results of open hypothesis questions. Note that dynamic task generation systems, e.g., Crowd4U~\cite{crowd4u}, can also be designed. The cross-tabulations are updated as each task is performed. Thus, when we stop submitting tasks, we obtain the best-effort results at any time. 

\subsubsection*{Limiting the Number of Closed Hypothesis Questions}
Although computing the exact ranking requires many closed hypothesis questions, there are two opportunities to reduce their number so that they can be considered as microtasks. 
\begin{description}
\item[Principle 1] Focus on hypotheses that have higher odds ratios. 
\item[Principle 2] Focus on hypotheses that are representative of similar ones.
\end{description}

The two principles are implemented for choosing a fixed number of hypotheses from the closed hypothesis question task. The inputs are twofold: 1) a set $H_t$ of hypotheses that have already been obtained at time $t$ in Phase 1; 2) a natural number $m>0$. The algorithm then outputs $H_{Q_{closed}}$ such that $H_{Q_{closed}} \subset H_t$ and $|H_{Q_{closed}}|=m$. According to Principle 1, a hypothesis is taken from $H_t$ and added to $H_{Q_{closed}}$ under Principle 2 while $|H_{Q_{closed}}|<m$ and $H_t$ is not a null set.

\subsection*{Phase 2: Verifying Hypotheses}
The output of Phase 1, which is computed based on workers' experiences so far, is often unsuitable, including a pseudo-correlation in the opposite direction of causality. To verify hypotheses at a higher evidence level~\cite{ustaskforce}, we employed an automatic pseudo-RCT, in which subjects were randomly divided into two groups. To balance the number of hypotheses attempted to the number of those not attempted, the subjects were divided into two groups, with one group attempting the hypotheses in the first half of the trial and the other in the second half. 

Phase 2 consists of three tasks, including two trials. Each trial employs subject-recruitment and result-reporting tasks.

\begin{description}
    \item[Subject-recruitment tasks] The tasks provide instructions to workers who will perform offline actions to validate the hypothesis hi. This task aims to answer $Q_{outcome}$, i.e., determine whether the worker is in the condition and follows the instructions for offline actions that are a possible cause of the outcome.
    \item[Result-reporting tasks] Workers are asked to report the effects of their offline actions. Instead of asking them to explicitly report the results, it poses $Q_{outcome}$ again to determine their current condition. These tasks are only open to workers who performed the subject-recruitment tasks.
\end{description}

Given a hypothesis $h_i$, let $R_i^1$, $R_i^2$, and $R_i^3$ be the results of Tasks 1, 2, and 3, respectively. Our framework examines whether there is a significant difference between the results of the hypotheses considered and those not considered. For example, the difference between $R_i^1$ and $R_i^2$ is regarded as the result of hypotheses that have been considered, whereas the difference between $R_i^2$ and $R_i^3$ is regarded as the result of hypotheses that have not been considered by the group that attempted the hypotheses in the first half. 

\subsection*{Experiments}  
\subsubsection*{Settings}
Experiments on sleep quality improvement were conducted to evaluate the proposed framework. 
For the experiments, we chose the topic ``getting a good night's sleep" as the outcome because this topic is familiar to everyone. To evaluate the degree of good sleep, we utilized a questionnaire, PSQI~\cite{psqi}, which is a popular indicator for measuring sleep quality. 

The tasks were created on the Crowd4U~\cite{crowd4u} platform, and workers were recruited from Yahoo! Crowdsourcing~\cite{yahoo}, which is a major crowdsourcing service in Japan, where people with various attributes are indiscriminately included as crowd workers. The tasks were created in Japanese, and all answers were received in Japanese.

\subsubsection*{Hypothesis Generation and Ranking Tasks}
\begin{figure}[!t]
 \centering
 \includegraphics[clip,width=10cm]{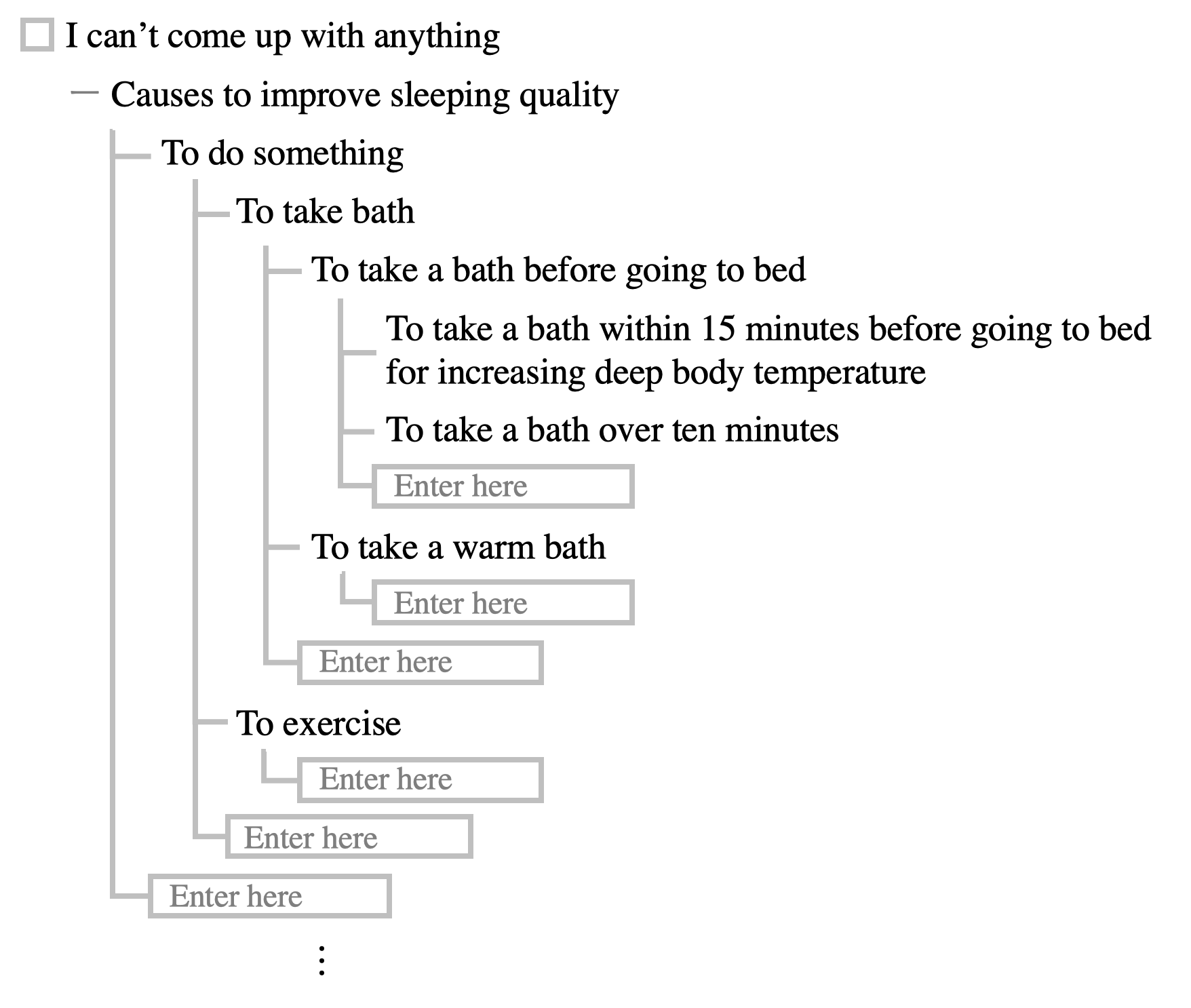}
 \caption{The tree structure of the open hypothesis question. Hypotheses are shown as tree nodes. This structure enables workers to add a new hypothesis by looking at generic and specific hypotheses.}
 \label{fig3}
\end{figure}

Given ``getting a good night's sleep" as the outcome, the outcome question $Q_{outcome}$ was set to the simple question ``Do you sleep well?" followed by PSQI questions. Next, the open hypothesis question $Q_{open}$ was set to ``What do you think is a possible cause of good sleep?" and associated with the list of hypotheses already given by other workers. The list was presented in a tree structure (Fig~\ref{fig3}), with each hypothesis corresponding to an individual tree node, such as intermediate nodes and leaf nodes. The entry fields are placed as new leaf nodes, allowing workers to add new hypotheses that share the same ancestors as their siblings. Therefore, the hypotheses are classified naturally as soon as they are added to the tree. In the case that workers had no ideas, they could skip this process without adding any hypotheses. For the closed hypothesis question $Q_{closed}$, the question number $m$ was set to ten. Note that these ten hypotheses were selected from the leaf nodes of the tree structure. To prevent meaningless spam input from affecting the process, the workers could choose ``the question does not make sense" in answering $Q_{closed}$ if they felt the hypothesis was not relevant. We offered 3 JPY (about 0.03 USD) to each worker who performed the task and left the tasks open for ten days. 

Principles 1 and 2 for this experiment were implemented as follows. 
\begin{description}
\item[Principle 1 Implementation] A hypothesis was randomly chosen according to the odds ratio-based weight $P(h_i)$ of each hypothesis $h_i$ in $H_t$. This is based on the observation that hypotheses with a higher odds ratio at a certain point are likely to have a higher odds ratio in the final state. As the number of hypotheses grows, we encounter the cold start problem: newly obtained hypotheses have very small $P(h_i)$ values and rarely have a chance to obtain higher odds ratios by appearing in tasks. To avoid this, we used the maximum value at that time for $P(h_i)$ until ten answers had been obtained, so that these hypotheses would be selected with priority.
\item[Principle 2 Implementation] To determine whether a similar hypothesis already existed in $H_{Q_{closed}}$, we used Fisher's exact test and measured a correlation between the sets of people who experienced two hypotheses. If there is a hypothesis $h'$ in $H_{Q_{closed}}$ that is correlated with hi in terms of the sets of people who experienced $h_i$ and $h'$, then this function returns ``true," and we do not add hi to $H_{Q_{closed}}$ because $h'$ is representative of similar hypotheses such as $h_i$.
\end{description}

\begin{figure}[!t]
 \centering
 \includegraphics[clip,width=12cm]{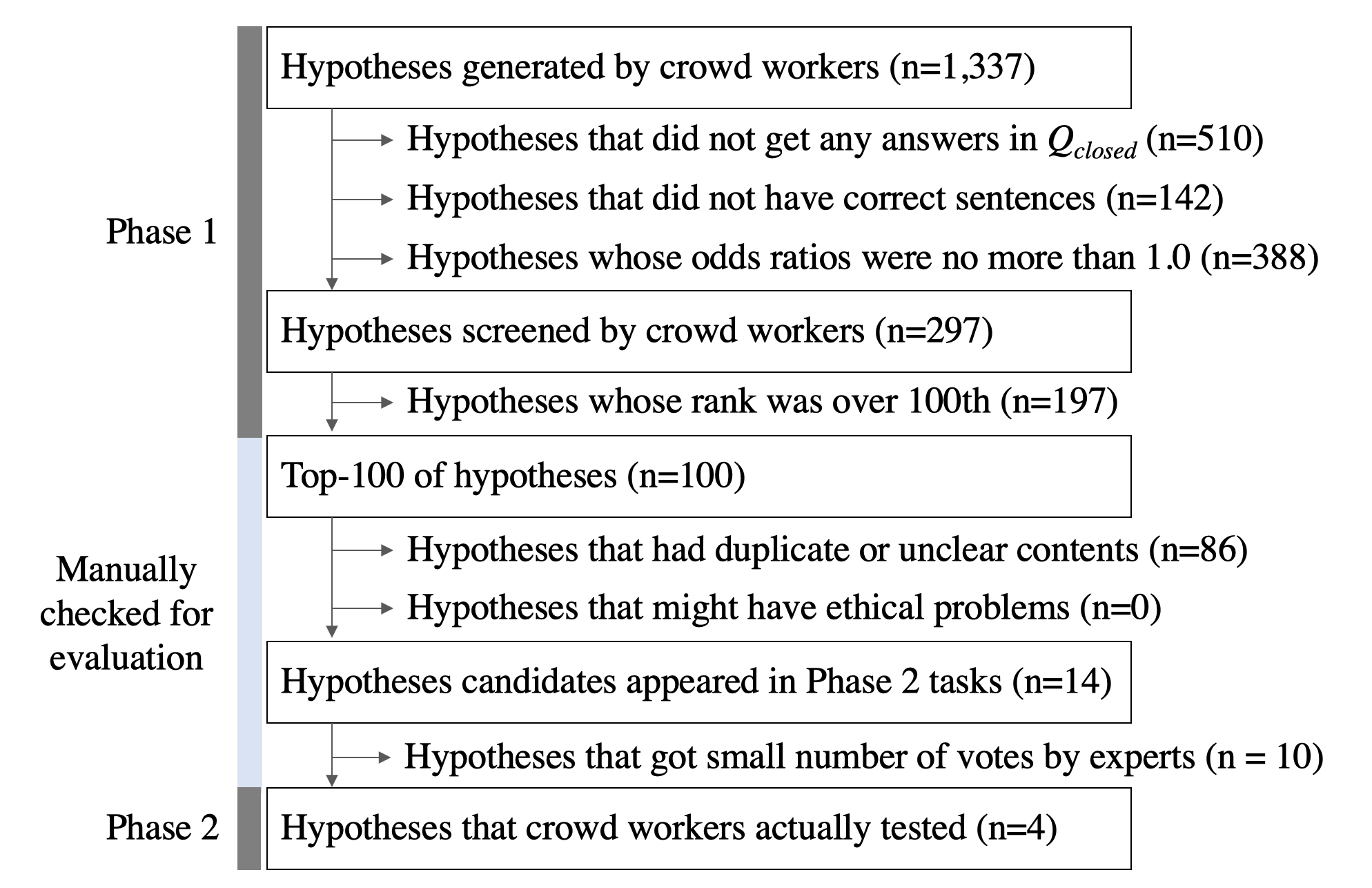}
 \caption{The procedure of hypotheses selection in the experiments. The upstream part was conducted by crowd workers in Phase 1. The downstream part was conducted by the authors and sleep researchers for the evaluation in Phase 2.}
 \label{fig4}
\end{figure}

To increase the number of workers assigned to each hypothesis, we decided to verify a smaller number of hypotheses in the experiment, and we thus formulated an appropriate procedure. Note that although this procedure was specific to the experiments described in this paper, our framework can be used for other approaches.

Fig~\ref{fig4} shows the entire procedure for selecting the hypotheses. The upstream part was conducted by crowd workers in Phase 1. In the downstream part, we manually selected hypotheses from the top 100 that did not include ambiguous or duplicate expressions. Then, we asked eight sleep researchers to exclude ethically problematic hypotheses and add one of three labels to each hypothesis: ``seems to be effective," ``seems to be ineffective," or ``neither agree nor disagree.'' Finally, we chose hypotheses for the Phase 2 tasks in the descending order of the number of votes for each label. Note that this intervention by people who were not recruited through crowdsourcing is not an essential part of our framework; it was necessary in this instance to pass the research ethics review and obtain data for the evaluation. In fact, the first two steps could be performed by non-expert crowd workers, and it is not necessary the labeling by experts that was done to limit the number of hypotheses in the Phase 2 tasks and save time and monetary cost.

A total of 1,337 hypotheses were collected in Phase 1, including those that correspond to intermediate nodes of the tree in $Q_{closed}$. We regarded votes for a leaf node in $Q_{closed}$ as those for intermediate nodes on the path from the leaf to the root. The odds ratios of the hypotheses associated with those nodes were computed using the number of votes. As a result, the four hypotheses were selected as shown in Table~\ref{table2}.

\begin{table}
\centering
\caption{Four hypotheses that were selected for the Phase 2 tasks.}
\begin{tabular}{rll} \toprule
ID & Hypothesis & Experts' answer\\ \midrule
1 & Get up early and bask in the sun & Seems to be effective\\
2 & Sleep sideways with the heart down & Neither agree nor disagree\\ 
3 & Make own eyes tired using smartphones & Seems to be ineffective\\
4 & Eat three meals a day by chewing well & Neither agree nor disagree\\
\bottomrule
\end{tabular}
\label{table2}
\end{table}

\subsubsection*{Hypothesis Verification Tasks}
\begin{figure}[!t]
 \centering
 \includegraphics[clip,width=12cm]{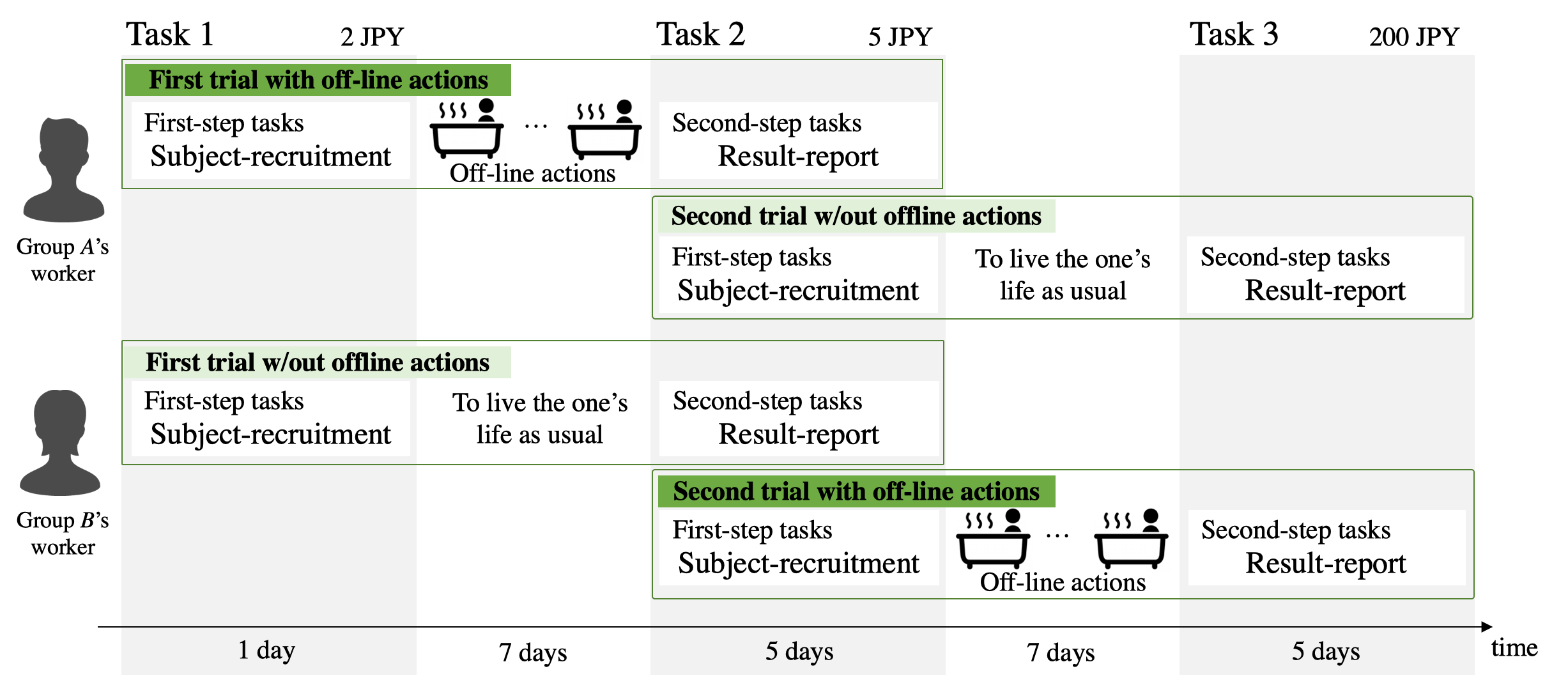}
 \caption{ Settings of Phase 2 tasks in the experiment.}
 \label{fig5}
\end{figure}

Fig~\ref{fig5} shows the setting of the Phase 2 tasks. For the hypothesis-verification tasks, the time interval between subject recruitment and reporting results was set to one week. Task 1 was open in the crowdsourcing service for a day. Tasks 2 and 3 were added seven days later and kept open for five days.

We paid 2 JPY (about 0.02 USD) to each worker who performed Task 1, 5 JPY (about 0.05 USD) to those who performed Task 2, and 200 JPY (about 2 USD) to those in Task 3. 

\section*{Results}
\subsection*{Hypothesis Generation and Ranking Tasks}

\begin{figure}[!t]
 \centering
 \includegraphics[clip,width=10cm]{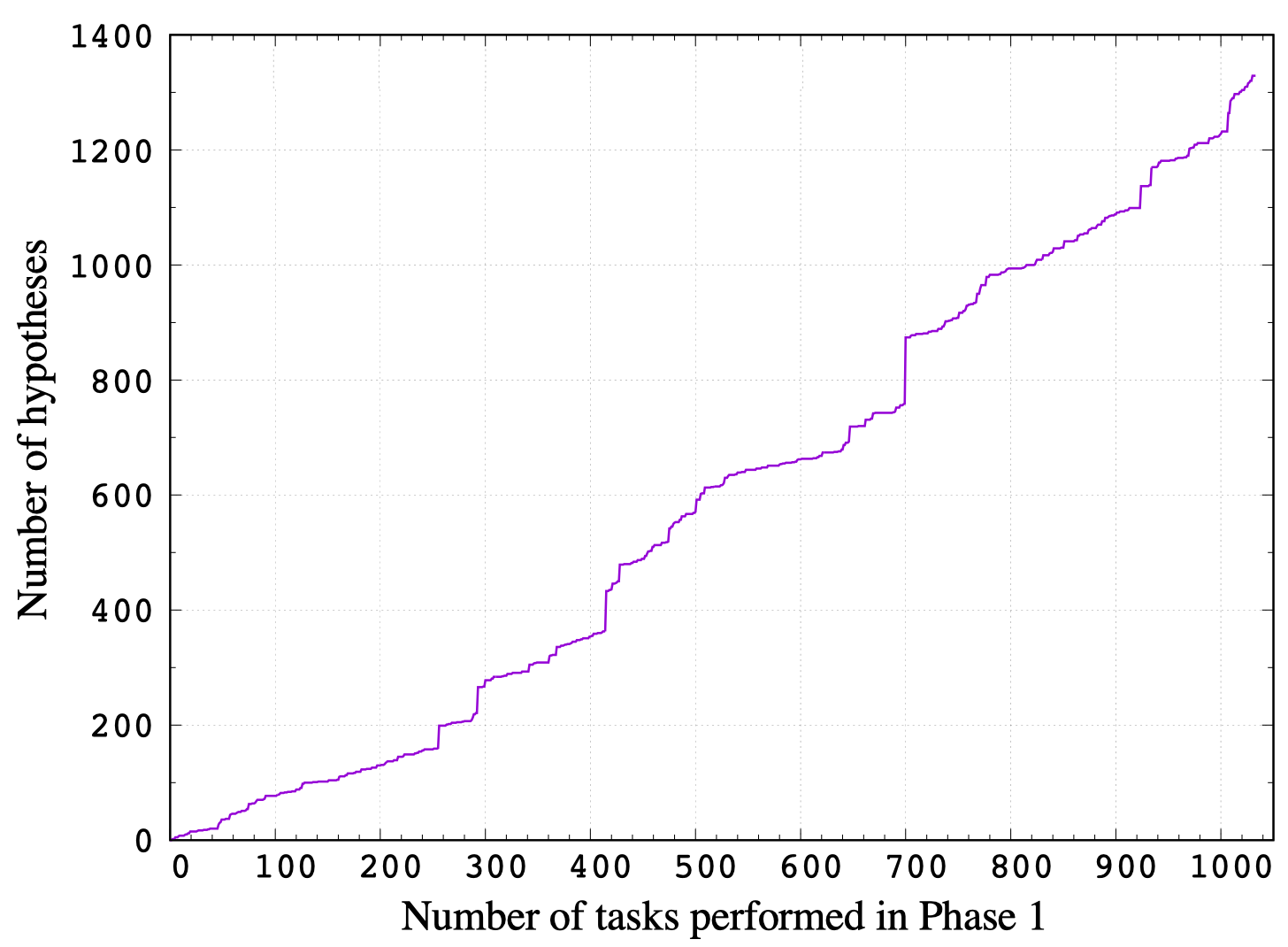}
 \caption{Settings of Phase 2 tasks in the experiment.
 }
 \label{fig6}
\end{figure}

In Phase 1, 1,031 tasks were performed, and 10,100 JPY (92 USD) was paid to crowd workers. A total of 1,337 hypotheses were generated in the hypothesis generation stage. The number of hypotheses increases linearly with the number of tasks; workers constantly enter new hypotheses that are slightly different from existing hypotheses (Fig~\ref{fig6}). For example, if there is a hypothesis like ``Take a bath before going to bed," someone may suggest ``Take a bath 10 minutes before going to bed." This suggests that there is no easy way of determining when to stop collecting the hypotheses and start ranking them. Thus, pipeline execution is required.

Overall, 827 hypotheses received one or more answers for $Q_{closed}$. After excluding hypotheses for which two or more workers answered, ``the question does not make sense," 685 hypotheses remained. We manually examined the top-ranked 100 hypotheses and did not find any obviously meaningless entries. This suggests that the ranking method in which users can answer ``the question does not make sense" in $Q_{closed}$ works well in reality. The odds ratios of 297 out of the 685 hypotheses exceeded 1.0, whereas 312 hypotheses had values of less than 1.0. The maximum odds ratio was 63 and the minimum was 0.022.

\begin{figure}[!t]
 \centering
 \includegraphics[clip,width=12cm]{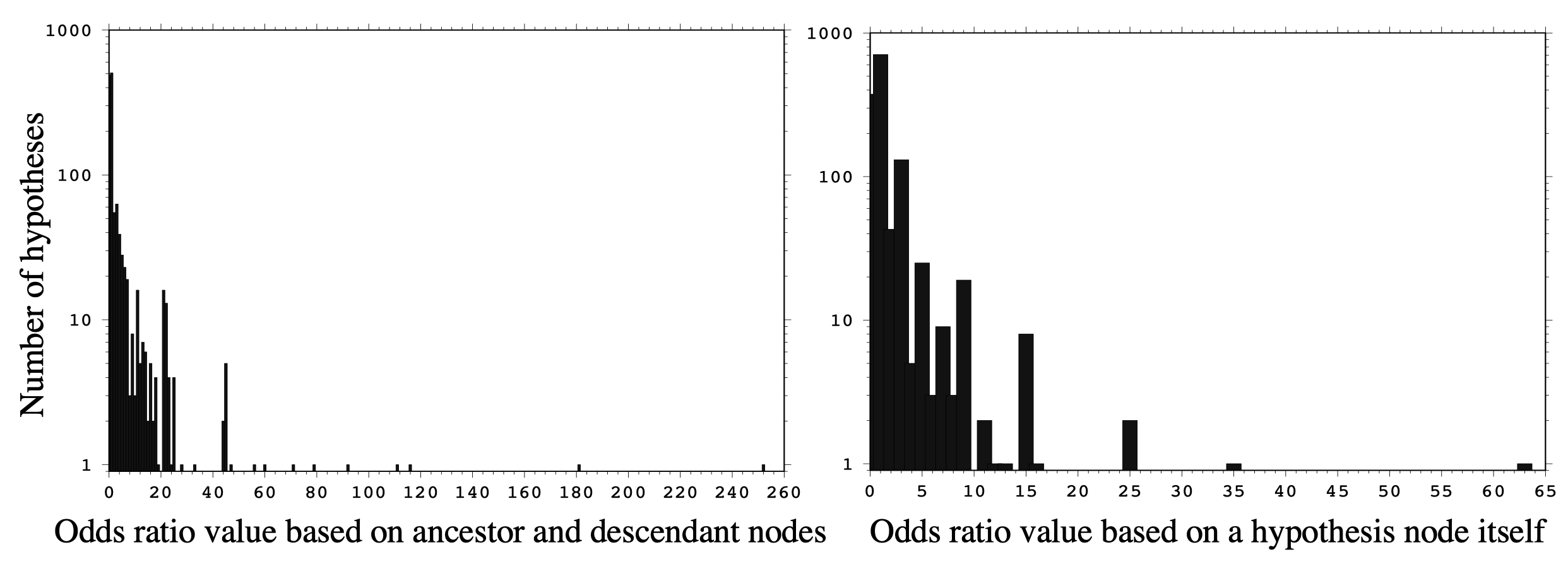}
 \caption{Frequency distribution of hypotheses. Many insignificant hypotheses were generated by workers in the Phase 1 tasks. The left graph shows the odds ratios calculated by the values of the ancestor and descendant nodes when hypotheses are arranged in a tree structure. The right graph shows the odds ratios calculated by the values of the hypothesis itself.
 }
 \label{fig7}
\end{figure}

Fig~\ref{fig7} shows the frequency distribution of the hypothesis odds ratios. The left graph presents the odds ratios calculated using the ancestor and descendant node values, while the right graph shows the odds ratios calculated using the value of the node. We can see that aggregating votes to compute the odds ratios for hypotheses that correspond to internal tree nodes effectively generate hypotheses with a larger number of votes. In addition, both results indicate that few significant hypotheses were generated by workers in Phase 1. 

\subsubsection*{Hypothesis Verification Tasks}
The hypotheses used in the hypothesis verification tasks are shown in Table~\ref{table2}. 1,153 workers responded to Task 1, 550 workers responded to Task 2, and 469 workers responded to Task 3. Therefore, approximately half of the Task 1 participants continued to Tasks 2 and 3. We paid 8,000 JPY (73 USD), 7,692 JPY (70 USD), and 139,071 JPY (1,262 USD) for Tasks 1, 2, and 3, respectively.

\begin{figure}[!t]
 \centering
 \includegraphics[clip,width=10cm]{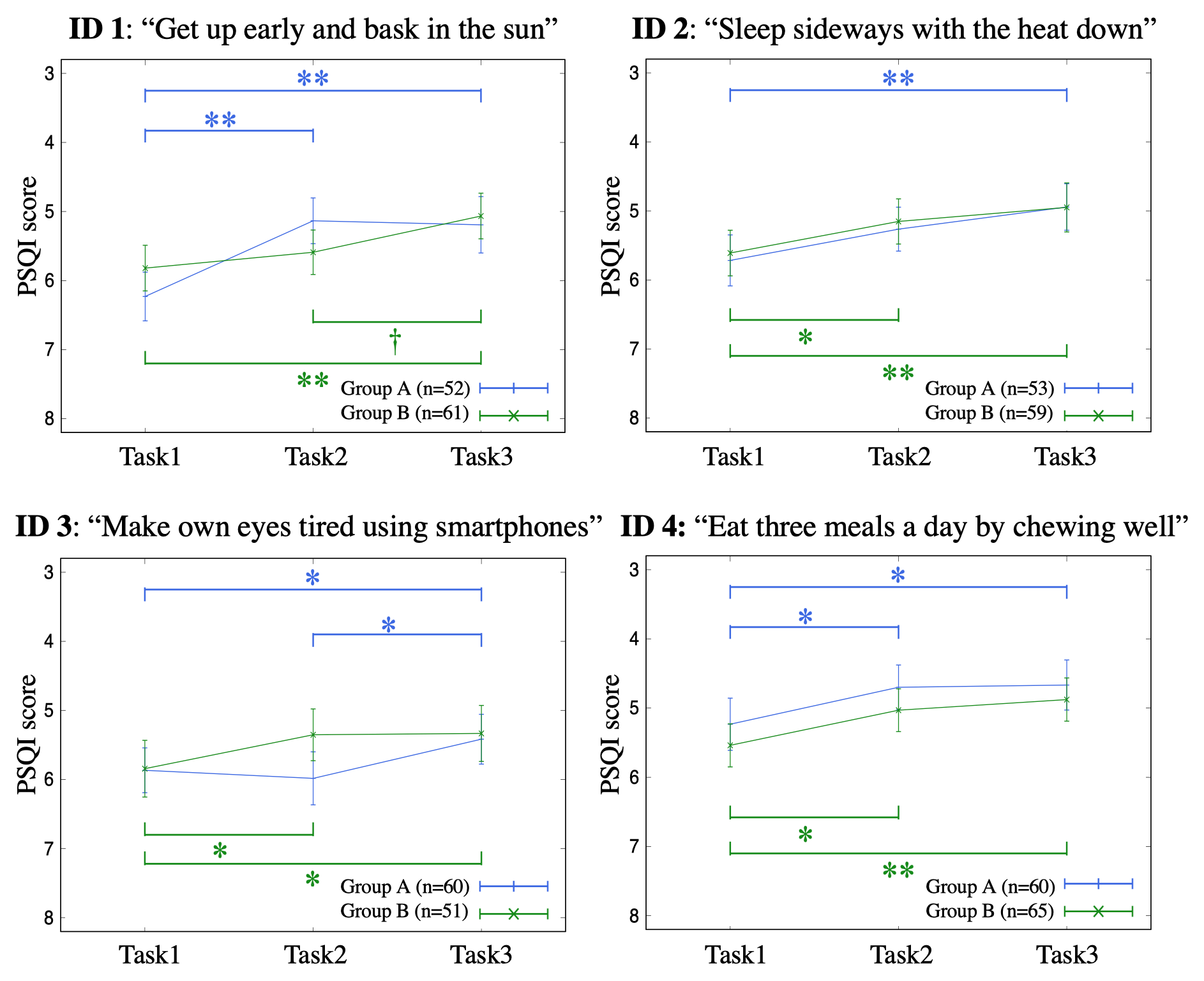}
 \caption{The blue/green lines represent Group A/B (tested the hypothesis in the first/second trial), respectively. The lower the PSQI score, the better the quality of sleep. $n$ is the number of workers who performed all three tasks (\dag: $p<.10$, \*: $p<.05$, \*\*: $p<.01$).
 }
 \label{fig8}
\end{figure}

Fig~\ref{fig8} shows the transition of the mean PSQI scores for each group of workers under each hypothesis. A paired t-test was conducted on the difference in PSQI scores among the three tasks. Overall, significant differences in PSQI scores were observed between Tasks 1 and 3 under all four hypotheses. This might be the effect of self-selection bias. According to the comparison between Tasks 1 and 2, or Tasks 2 and 3, for hypothesis 1, significant differences were only observed in the trial when this hypothesis was tested. In contrast, for hypothesis 3, significant differences were only observed in the trial when this hypothesis was not tested. 

\begin{figure}[!t]
 \centering
 \includegraphics[clip,width=10cm]{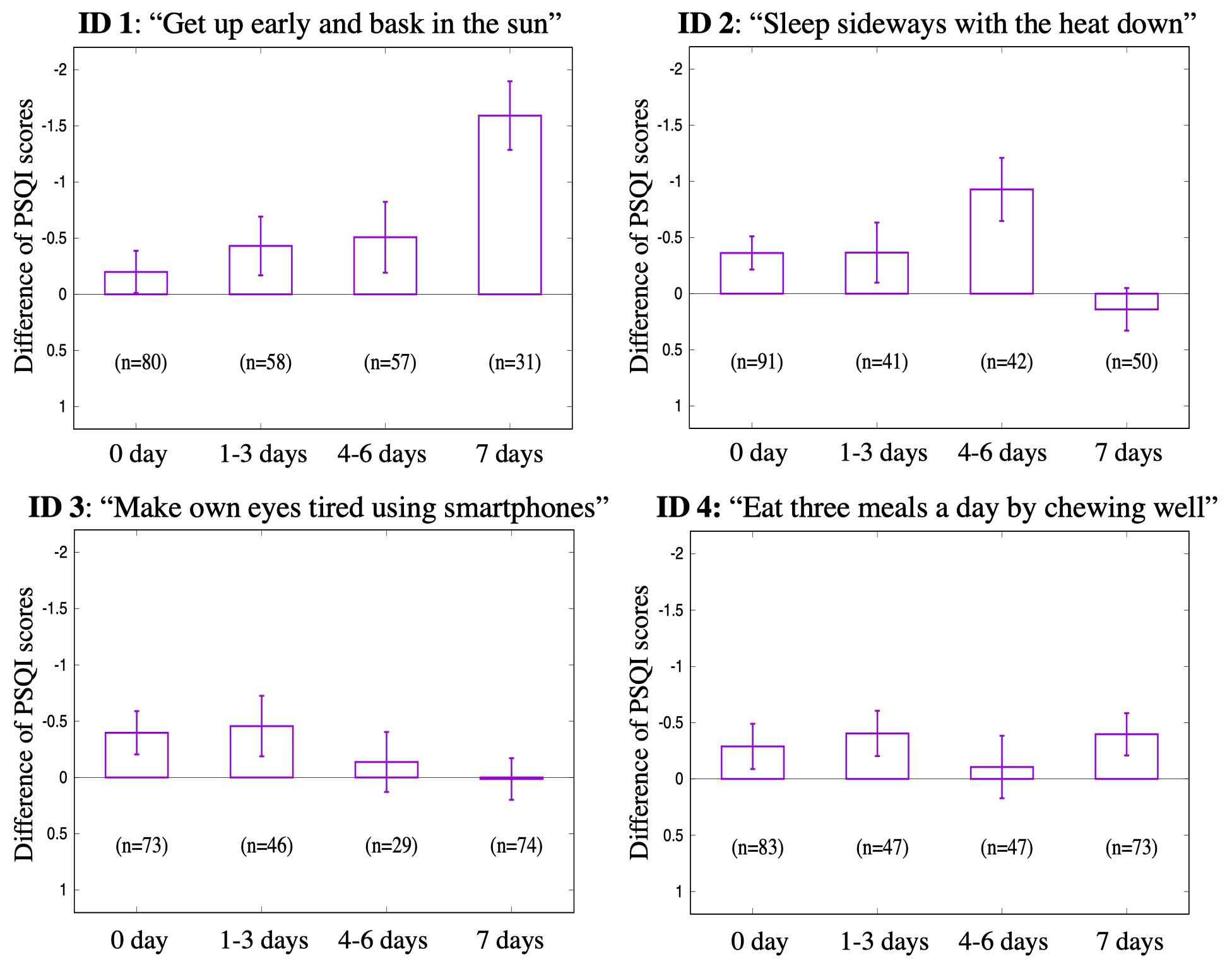}
 \caption{Differences in mean PSQI score for each number of days the workers tested the hypothesis within a week (with standard error bars). Y-axis shows the difference between the PSQI scores before and after the trial.
 }
 \label{fig9}
\end{figure}

Fig~\ref{fig9} shows the relationship between how many days the workers actually tested the assigned offline action and the difference in PSQI score. As the number of days increased, the PSQI scores improved for hypothesis 1. In contrast, for hypothesis 3, the PSQI scores became worse as the number of days increased. Under hypothesis 2, the PSQI scores improved and worsened, and there were no notable changes for hypothesis 4.

For hypothesis 1, the mean PSQI scores only exhibited significant improvement in the week when the hypothesis was tested. From this, hypothesis 1 can be considered an effective cause of improved sleep quality. This conclusion is consistent with the majority vote of the experts (see Table~\ref{table2}). In contrast, for hypothesis 3, the mean PSQI scores only significantly improved in the week when the hypothesis was not tested. That is, sleep quality only improved when the workers testing hypothesis 3 stopped performing the associated actions, suggesting this is counterproductive for improving sleep quality. This conclusion does not contradict the majority vote of the experts in Table~\ref{table2}. For hypotheses 2 and 4, there were no significant differences between when these hypotheses were and were not being tested. These hypotheses were also categorized as ``Neither agree nor disagree" by the experts (see Table~\ref{table2}).

The above results suggest that hypothesis verification by workers in our framework produces the same outcomes as that by the experts. Note that we did not use a special mechanism to exclude spam in Phase 2. However, the results suggest that if $N$ is sufficiently large, the results will converge if the hypothesis does not refer to a placebo effect. This is not surprising because it reflects the principle of statistics. We can exclude the effects of noise if we have a large number of trials.

The number of workers who performed tasks in Phase 2 was more than that in Phase 1. From this result, it can be considered that the number of workers for each phase varies widely depending on the seasons or rewards. Although there were three tasks in Phase 2 in the experiments, about 60\% of the workers dropped out before Task 3, which also caused a sampling bias. Thus, we obtained important findings through the Phase 2's experiments.

\section*{Discussion}
\subsection*{Principal Results}
Experimental results demonstrated that the proposed framework could collect various hypotheses from the crowd. One of the advantages of our framework is that we could investigate hypotheses focusing only on feasible ones because the workflow was clearly divided into the two phases: Phase 1 for generating and ranking hypotheses, and Phase 2 for verifying hypotheses. More than 1,100 people participated per day, and about 40\% of them performed tasks until the last with pay of only 207 JPY per worker. In total, we paid 164,863 JPY (1,500 USD) to generate multiple hypotheses and verify four of them. Although the proposed framework cannot replace the quality of an experts' work, it is much cheaper than experiments conducted by experts. 

Consequently, the proposed framework works well for a certain class of research questions; although we could not verify whether the workers really performed the assigned task, the crowd verification produced different results from those drawn from observation only. In fact, the test results with a large number of workers (e.g., $N>100$) converged to a conclusion that the experts can agree on, even if a certain number of workers did not behave as expected. The results clearly show that this approach stands between two extreme solutions in effect. 

\subsection*{Limitations}
\subsubsection*{A Micro Research Framework}
We need to be careful in designing the workflow in the hypothesis generation and ranking tasks. Since we found that we would have many duplicate hypotheses in our preliminary experiment, we showed workers a tree-like summarization of the hypotheses provided so far to avoid duplicates. Nevertheless, workers can generate an infinite number of hypotheses, including equivalent ones with different expressions. This leads to two problems: (1) It is difficult to know when to stop collecting the hypotheses and start selecting those to be verified; and (2) If we deal with each of the hypotheses separately, we cannot collect a sufficient number of answers to estimate their importance or a sufficient number of workers to verify them.

Although we found that the pipeline execution of questions and a tree-based aggregation of answers are effective, another problem is that crowd workers do not necessarily have loyalty. As our results suggest, workers gradually quit during a sequence of verification tasks. Therefore, we need to carefully design the procedure to balance the numbers of workers in the experimental and control groups.

In the hypothesis verification task, our method still has much room for improvement. First, it is not studied on the incentive that a crowd worker is actively implementing an intervention in their everyday life for a small economic return. Further, there is no evidence that a worker really completed their intervention. Nevertheless, the results showed the feasibility, suggesting that each micro result may not be not reliable, but the total aggregating results are reliable. One of the reasons is that the incentive could also be the participation in a study; however, a more detailed analysis of the sustainability of the economic approach could help to strengthen the motivations. 

Another remaining task is confounding. Because the crowdsourcing approach is not controlled, it is easily biased by confounding variables. A more controlled setting, such as a within-subjects design and a between-group design, is desirable. In addition, we cannot introduce placebos. Thus, the results here are biased by the participant effect. Removing such biases is one of our interesting future work. When workers vote for others' ideas in crowdsourcing, it might be better to set weights~\cite{hardas2012}. There are many other parameters we can explore for better designs.

\subsubsection*{Applicability Conditions}
According to our assumptions regarding crowd workers, the micro research works on problems that satisfy the following two conditions.
\begin{itemize}
    \item Given a specific phenomenon, there must be two types of people in the worker set: people who experience the phenomenon (e.g., good sleepers) and those who do not (e.g., bad sleepers). 
    \item The hypotheses should be related to things workers can observe, including the environment around them, their experiences, actions, and physical conditions. Therefore, for example, hypotheses involving genetic disorders cannot be verified.
\end{itemize}

\subsubsection*{Evidence Level}
To find the real causal relations associated with a given phenomenon, we need to conduct designed experiments, in which we have total control over all variables. In contrast, the research method in which we draw inferences from a sample of a population where the independent variable is not under the control of the researcher is called an observational study.

Unfortunately, we cannot perform designed experiments in a crowdsourcing setting. For example, we cannot perform random sampling to assign crowd workers to the experimental group (the group of people on which the experimental procedure is performed) and the control group (the group of people on which the experimental procedure is not performed). This is one of the limitations of this study.

\subsection*{Related Work}
Traditional data quality management in crowdsourcing applications assumes that high-quality data are those received from reliable workers. Therefore, the idea behind many data aggregation methods is to assign higher weights to the answers given by workers who are likely to be reliable~\cite{berend2014,daniel2018,law2011,raykar2010}. 
In contrast, our framework is designed to find high-quality answers to research questions from a large number of workers regardless of their reliability. 

There have been many attempts at citizen science involving crowdsourcing~\cite{franzoni2014, nov2011,ramine2010,vaish2010}. For instance, Fold It~\cite{foldit} asks crowd workers to play an online puzzle game about protein structures for protein folding, whereas Galaxy Zoo~\cite{galaxyzoo} asks crowd workers to classify large numbers of galaxies. There are some attempts to perform a systematic review of research literature using crowdsourcing~\cite{Brown2014,wallace2017}
These projects exploit human computation capabilities to solve problems that cannot easily be solved by algorithms only. We use human computation and offline task capabilities to generate hypotheses and verify them, while controlling the tasks performed by the crowd. The overall effect serves as a scientist. 

In medical research, crowdsourcing has mainly been used to collect data concerning people~\cite{leiter2014}. Swan surveyed crowdsourcing approaches for the collection of data on people's health~\cite{swan2012}. The data collected are often used to construct prediction models~\cite{bongard2013}.
On the other hand, there have been attempts to obtain data from the crowd to help hypotheses on health. Bevelander et al.~\cite{bevelander2014} collected information on people's childhoods to understand obesity, while Aramaki et al.~\cite{aramaki2017} explicitly asked the crowd to provide hypotheses. In those attempts, experts were required to form hypotheses or choose some of them, and to verify the obtained hypotheses. There are online medical diagnosis services where people can ask medical questions to others, including ordinary people~\cite{crowdmed}. In such services, the answers are ranked in some way, but the rankings are not verified.

In Phase 1 of our proposed framework, we select a small number of hypotheses that are likely to be true in order to create questions for microtasks. Unlike the top-k selection of objects with complete data~\cite{alfaro2017}, we must select hypotheses under the condition that we do not know their actual probabilities of being true. This is related to the multi-armed bandit problem~\cite{auer2002}. However, our setting is different from a typical multi-armed bandit, as we choose more than one arm in each task (our task in Phase 1 asks workers to choose more than one hypothesis from a list), and new machines are dynamically added. There are various techniques for determining when the collection of data should be stopped~\cite{franklin2010}. Our problem is unique in terms of dealing with a variety of expressions to state the same or similar hypotheses, and thus, the number of collected data items increases linearly without sophisticated support. Although we use the stability of the top-ranked hypotheses at present, it would be interesting in the future to apply techniques proposed in other frameworks.

\section*{Conclusions}
This paper described the potential and limitations of the micro research framework, whereby crowd workers generate and verify hypotheses on a particular type of research question. Our experiments with real-world workers revealed that prospective crowdsourced studies are feasible if certain conditions are satisfied, although the workflows must be carefully designed.

In future work, we will try to find better ways of verifying hypotheses to produce higher evidence levels based on the observations made in our experiments; in particular, methods for organizing the groups will be a key issue. 

\section*{Acknowledgments}
This work was supported by JST CREST Grant Number JPMJCR16E3, Japan.

\end{document}